\UseRawInputEncoding
\documentclass[reqno,12pt]{article}
\usepackage{amsmath,amsfonts,amssymb,amsthm,amstext,amscd,eucal,xcolor}
\usepackage[all]{xy}
\usepackage{hyperref}
\usepackage{epsfig}
\usepackage{color}
\usepackage{graphicx}
\usepackage[active]{srcltx}
\makeatletter \@addtoreset{equation}{section}

\makeatletter\renewcommand\section{\@startsection {section}{1}{\z@}%
	{-3.5ex \@plus -1ex \@minus -.2ex}
	{2.3ex \@plus.2ex}%
	{\normalfont\large\bfseries}}
\renewcommand\subsection{\@startsection{subsection}{2}{\z@}%
	{-3.25ex\@plus -1ex \@minus -.2ex}%
	{1.5ex \@plus .2ex}%
	{\normalfont\bfseries}}

\newcommand{\be}{\begin{equation}}
	\newcommand{\ee}{\end{equation}}
\newcommand{\bea}{\begin{eqnarray}}
	\newcommand{\eea}{\end{eqnarray}}
\newcommand{\bse}{\begin{subequations}}
	\newcommand{\ese}{\end{subequations}}
\newcommand{\bi}{\begin{itemize}}
	\newcommand{\ei}{\end{itemize}}

\newcommand{\beq}{\begin{eqnarray}}
	\newcommand{\eeq}{\end{eqnarray}}

\newcommand{\nn}{\nonumber}

\def\s2s1{S$^2\times$S$^1$ }
\def\Label#1{\label{#1}%
	\smash{\hbox to0pt{\raise1ex\hbox{\tiny[#1]}\hss}}}
\def\noLabels{\let\Label=\label}
\def\nobbibitem{\let\bbibitem=\bibitem}

\begin{document}
	\baselineskip 18pt%
	\begin{titlepage}
		\vspace*{20mm}
		\begin{center}
			{\Large{\textbf{Regular Black Holes:\\
						Entropy Products and Central Charges}}}
			\vspace*{8mm}
			
			Hanif Golchin\footnote{h.golchin@uk.ac.ir} \\
			\vspace*{0.4cm}
			{ \it Faculty of Physics, Shahid Bahonar University of Kerman, \\
				PO Box 76175, Kerman, Iran} and \\
			{\it School of Physics, Institute for Research in Fundamental Sciences (IPM), P.O. Box 19395-5531, Tehran, Iran}
			\vspace*{1.5cm}
		\end{center}

\begin{abstract}
In this paper for variety types of regular black hole solutions, we investigate the entropy product of inner and outer horizons. Similar to singular black holes, for the regular ones we find that  universality (mass independence) of the entropy product is true for some solutions and it fails for some others. In the case of regular black holes that respect the universality, we read central charges of the dual CFTs from the entropy product, according to the Thermodynamics method introduced in \cite{Chen:2013rb}. For these solutions we also calculate central charges, using the asymptotic symmetry group formalism. The results of these two approaches are the same, which means that universality of the entropy product provides a simple method to find central charges of the dual CFTs. 
\end{abstract}


\end{titlepage}

\addtocontents{toc}{\protect\setcounter{tocdepth}{2}}
\tableofcontents

\section{Introduction}
According to the singularity theorem \cite{Hawking:1973uf}, formation of singularities are unavoidable (under some circumstances) in general relativity. The existence of singularity is the failure of general relativity and to protect it, Penrose proposed that singularities are covered by horizons (the weak cosmic censorship conjecture). It is believed that singularity appears in the classical theory of gravity and by taking into account the quantum effects, they can be avoided \cite{Sakharov:1966aja,Bojowald:2007ky}. Inspired by this idea, Bardeen proposed \cite{Bard-reg} a regular black hole (RBH) solution in which the singularity is replaced by a de Sitter core. After the seminal work of Bardeen, a large number of RBHs are constructed \cite{Ayon-Beato:1998hmi}-\cite{Roupas:2022gee}. Recently it is also proposed some  models for RBHs with a Minkowski core \cite{Ghosh:2014pba,Simpson:2021zfl,Simpson:2019mud,Simpson:2021dyo}. 

Black holes (BH)s are good locations to explore the relation of gravity and quantum mechanics.
It has been observed \cite{Cvetic:2010mn}-\cite{MahdavianYekta:2020lde} that the horizons entropy product for many BH solutions is universal (mass independent) as $S_+S_-=4\pi^2 N$, where $N$ is related to the quantized a charges of the solution, like angular momentum and electric/magnetic charge. It is shown \cite{Chen:2012mh} that the condition $T_+S_+=T_-S_-$, where $T_+$ ($T_-$) is the Hawking temperature of the outer (inner) horizon, is equivalent to the mass independence of $S_+S_-$ . This condition also implies that central charges of the left-moving and right-moving sectors in the dual conformal field theory (CFT) are the same ($c_L=c_R$). It is discussed in \cite{Chen:2013rb} that for the solutions with universal entropy product, one can read the central charge(s) of the dual CFT(s) as 
\be \label{unicc}
c_i=\frac{6}{4\pi^2}\,\frac{\partial \left(S_+S_-\right)}{\partial N_i}\,,
\ee
which is the Thermodynamics method for finding the central charge. 

There are also BH solutions in which the entropy product is not universal. For instance, it is observed \cite{Castro:2013pqa} that  the universality of entropy product fails for some BH solutions of higher curvature gravity. The entropy product is also mass dependent for the BHs  with NUT charge \cite{Xu:2015mna,Debnath:2015tda}. Although in the case of 5D Myers-Perry and the BTZ black hole the universality is true, it fails for the Myers-Perry BHs in $D\geq 6$ and for Kerr-AdS BHs in $D\geq 4$ \cite{Chen:2012mh}. 

In the case of RBHs it is shown \cite{Pradhan:2015wnl} that the universality fails for the ABG (Ay\'{o}n-Beato and Garc\'{i}a) BH. In \cite{Fan:2016hvf} it is shown that for RBHs of Einstein gravity coupled to extended nonlinear electrodynamics, by choosing adequate parameters $S_+S_-$ can be mass independent. However for large number of RBHs (maybe because of the complex form of the metric), the entropy product is not studied yet. 

This paper is organized as follows. In section \ref{s2} we briefly review the Thermodynamics method for finding central charge(s) of the dual CFT(s). Then we investigate the entropy product for RBHs where the universality fails for some of them (solutions in section \ref{s3} ) and it works for some others (section \ref{s4}). In section \ref{s5} using Thermodynamics method, we find the central charges of the dual CFTs for RBHs with universal entropy product, studied in section \ref{s4}. We then recalculate the central charges using the asymptotic symmetry group formalism to check the result of Thermodynamics method. We conclude the paper with a discussion on the universality of entropy product and its importance in the BH solutions.

Throughout the paper we set the Newton's gravitational constant to unity, i.e. $G=1$.
\section{Universality of entropy product and dual CFTs} \label{s2}
In this section we review some BH solutions with universal entropy product. For these solutions, it is possible to find central charges of the dual CFTs, using the Thermodynamics method.

\vspace{2mm}
$\bullet$ \, {\bf The Kerr-Newman BH}\\ 
The Kerr-Newman BH \cite{Newman:1965my} is determined by three conserved charges $M$ (mass), $J$ (angular momentum) and $Q$ (electric charge). In \cite{Ansorg:2009yi} it has been shown  that the entropy product of inner and outer horizons for this solution is universal as
\be \label{unikn}
S_+S_-=\pi^2(4J^2+Q^4)\,.
\ee
The dual CFT for this solution is studied in \cite{Chow:2008dp,Chen:2011wm} where it is shown that the $J$-picture central charge $c^J$ and the $Q$-picture central charge $c^Q$ are give by
\be \label{cjcq}
c^J=12\,J\,, \qquad c^Q=6Q^3\,.
\ee
Due to the universality of entropy product, one can also find these central charges by using Thermodynamics method (\ref{unicc}) as
\bea
&&c^J=\frac32\,\frac{\partial}{\partial J}\,\Big(4J^2+Q^4\Big)=12J\,, \nn\\ &&c^Q=\frac32\,\frac{\partial}{\partial Q}\,\Big(4J^2+Q^4\Big)=6\,Q^3\,,
\eea
which are in agreement with (\ref{cjcq}).

\vspace{2mm}
$\bullet$ \, {\bf The Myers-Perry BH}\\
The Myers-Perry BH \cite{MP} in five dimensions is characterized by its two angular momenta $J_{\phi}$, $J_{\psi}$ and mass $M$. Applying the Kerr/CFT analysis, it is shown that \cite{Lu:2008jk, Ghodsi:2013soa, Ghodsi:2014fta} for the Myers-Perry BH, there are two dual CFT descriptions associated with the rotations along $\phi$ and $\psi$ directions. The central charges of these dual CFTs are 
\be \label{ccmp}
c_{\phi}=6J_\psi\,, \qquad c_{\psi}=6J_\phi\,.
\ee
It is also shown that \cite{Cvetic:2010mn,Chen:2012mh} the horizons entropy product for this solution is universal, {\it i.e.}
\be \label{ssmp}
S_+S_-=4\pi^2J_{\phi}J_{\psi}\,.
\ee
Now using Thermodynamics method (\ref{unicc}), one can easily find the central charges (\ref{ccmp}) from the above entropy product.


There are also more examples for the relation between entropy product and central charges that we refer the reader to \cite{Chen:2012mh,Chen:2013rb,Golchin:2020mkh,MahdavianYekta:2020lde}. These examples tell us that the universality of entropy product implies the existence of dual CFTs. In the following we check the validity of Thermodynamics method in the case of RBH solutions. To this end, we should find solutions with universal entropy product, so in the next sections we investigate the universality for RBHs.

\section{RBHs with non-universal entropy product} \label{s3}
As we mentioned, the product of horizon entropies is mass independent  if the condition $T_+S_+=T_-S_-$ is satisfied \cite{Chen:2012mh}. In this section and the next one, by explicit calculations we investigate this universality condition for RBHs. We consider variety types of regular solutions in four and higher dimensions, with/without angular momentum and charge. We also consider RBHs with both de Sitter and Minkowski cores. We find BH horizons by solving $g^{rr}=0$ which may have negative or complex roots. By outer (inner) horizon we mean the largest (smallest) positive real root.

\subsection{The Bardeen BH}
Bardeen BH is the first known RBH solution \cite{Bard-reg}. The matter source which supplies this solution was unknown for many years till Ay\'{o}n-Beato and Gars\'{i}a showed that the Bardeen BH is a magnetic solution of Einstein equations coupled to a nonlinear electrodynamics \cite{Ayon-Beato:1998hmi,Ayon-Beato:2000mjt}. The metric of Bardeen BH is in the form 
\bea \label{bardmetr}
&&ds^2=-f dt^2+\frac{dr^2}{f}+r^2 \left(d\theta^2+\sin^2 \theta d\phi^2\right),\\
&& \hspace{1.6cm} f=1-\frac{2M r^2}{(r^2+q^2)^{3/2}}\,,\nn
\eea
where $M$ is the mass and $q$ is the magnetic monopole charge. In Bardeen BH, the singularity at $r=0$ is replaced by a de Sitter core. For this solution, there are three horizons (roots of $f=0$) which are too long to be written here. We denote the outer and inner horizons by $r_+$  and $r_-$ respectively. The entropy and temperature of Bardeen BH on the outer and inner horizons are 
\be
S_\pm=\pi r_\pm ^2 \,,\qquad T_\pm=\pm \frac {M{r_\pm} \left( {{r_\pm}}^{2}-2{q}^{2} \right) }{2\pi\left( {q}^{2}+{{r_\pm}}^{2} \right) ^{5/2} }\,.
\ee
It is straightforward to check  that $T_+S_+\neq T_-S_-$, which means that entropy product is not universal for the Bardeen BH.

\subsection{The Hayward BH}
Another RBH solution with a de Sitter core is the Hayward BH \cite{Hayward:2005gi}. For this solution, the metric is in the static spherically symmetric form (\ref{bardmetr}) with the function
\be \label{fhay}
f=1-\frac{2M r^2}{r^3+q^3}\,.
\ee
Similar to the Bardeen BH, we find the entropy and temperature on the outer and inner horizons as 
\be
S_\pm=\pi r_\pm^2\,,\qquad T_\pm=\pm \frac{M r_\pm(r_\pm^3-2q^3)}{2\pi (r_\pm^3+q^3)}\,, 
\ee
one can check that the condition $T_+S_+=T_-S_-$ does not satisfied here, so the universality of entropy product is not valid for the Hayward BH.

\subsection{The rotating Bardeen and Hayward BHs}
Extension of the Bardeen and Hayward RBHs to the solutions containing angular momentum is done in \cite{Bambi:2013ufa}. The line element for these solutions is
\be \label{rotreg}
ds^2=-F(r,\theta)dt^2+\frac{dr^2}{G(r,\theta)}+\Sigma(r,\theta)d\theta^2
+H(r,\theta)d\phi^2-2K(r,\theta)dtd\phi\,,
\ee
with the functions
\bea 
F(r,\theta)&=&\frac{\Pi(r)-a^2\sin^2\theta}{\Sigma(r,\theta)}, \quad G(r,\theta)=\frac{\Pi(r)}{\Sigma(r,\theta)}\,, \quad
\Sigma(r,\theta)=r^2+a^2\cos^2\theta, \nn\\
 K(r,\theta)&=&\frac{2r m(r) a\sin^2\theta}{\Sigma(r,\theta)}\,,\qquad H(r,\theta)=\frac{(r^2+a^2)^2-\Pi(r) a^2\sin^2\theta}{\Sigma(r,\theta)}\,\sin^2\theta\,,   
\eea
the function $\Pi(r)$ in the above is introduced as $\Pi(r)=r^2+a^2-2r m(r)$ , where $a$ is the rotation parameter. In order to remove the singularity, $m(r)$ is given in the form
\be\label{massreg}
m(r)=M\left(\frac{r^p}{r^p+r_0^p}\right)^{3/p}\,.
\ee
The rotating Bardeen and Hayward BHs are constructed by setting $p=2$ and $p=3$ in (\ref{massreg}), respectively. Note that for $r>>r_0$ the above mass reaches the Kerr BH mass $M$. Moreover in the case of $r_0=0$ one finds $m(r)=M$ and (\ref{rotreg}) reduces to the Kerr solution, in other words, one can interpret $r_0$ as the deviation parameter from the Kerr BH. Horizons of the solution are roots of $\Pi(r)=0$. In order to find entropy and temperature on the inner and outer horizons, it is more convenient to solve $\Pi(r_+)=0$, $\Pi(r_-)=0$ and find $a$, $M$ in terms of $r_-$ and $r_+$\,. Noticing this point, we calculate entropy and temperature
for the rotating Hayward BH ($p=3$). The result is
	\bea
	S_+\!&=&\!\frac {\pi \,{ r_+}^{4} \left( r_-+ r_+ \right) 
		\left( { r_0}^{3}+{ r_-}^{3} \right) }{ \left(  r_-+r_+ \right)  \left( {r_-}^{2}+{r_+}^{2} \right) {r_0}^{3}+{r_-}^{3}{r_+}^{3}}\,, \qquad S_-=\frac {\pi \,{ r_-}^{4} \left( r_-+ r_+ \right) \left( { r_0}^{3}+{ r_+}^{3} \right) }{ \left(  r_-+r_+ \right)  \left( {r_-}^{2}+{r_+}^{2} \right) {r_0}^{3}+{r_-}^{3}{r_+}^{3}}\,,\nn\\
	T_+\!&=&\!\frac{(r_--r_+)\left[2r_0^6(r_++r_-)^2-r_0^3r_+^2(4r_-^3+4r_-^2r_++2r_-r_+^2+r_+^3)-r_-^3r_+^5\right]}{4\pi r_+^3(r_++r_-)(r_0^3+r_-^3)(r_0^3+r_+^3)}\,,\nn\\
	T_-\!&=&\!\frac{(r_--r_+)\left[2r_0^6(r_++r_-)^2-r_0^3r_-^2(4r_+^3+4r_+^2r_-+2r_+r_-^2+r_-^3)-r_+^3r_-^5\right]}{4\pi r_-^3(r_++r_-)(r_0^3+r_-^3)(r_0^3+r_+^3)}\,.
	\eea
In the case of rotating Bardeen BH ($p=2$), $S_\pm$ are messy and we do not show them here. We checked that for both rotating Bardeen and Hayward solutions $T_+S_+\neq T_-S_-$, means that the entropy product is not universal for them.  
In the case of small values for deviation parameter $r_0$, by keeping the first order in Taylor expansion around $r_0=0$, it is possible to find perturbed  (around the Kerr) solution and then calculate the entropy and temperature. For instance we find entropy and temperature for the Bardeen BH with small $r_0$ as
\bea
&&S_+=\frac{\pi r_-(r_-+r_+)(3r_0^2-2r_+^2)}{3r_0^2+2r_-r_+}\,,\nn\\
&&S_-=\frac{\pi r_+(r_-+r_+)(3r_0^2-2r_-^2)}{3r_0^2+2r_-r_+}\,,\nn\\
&&T_+=\frac{(r_+-r_-)\left[r_0^2(r_-+2r_+)+\frac{2}{3}r_-r_+^2\right]}{4\pi r_+r_-(r_++r_-)\left(r_0^2-\frac{2}{3} r_+^2\right)}\,,\nn \\ &&T_-=\frac{(r_+-r_-)\left[r_0^2(r_++2r_-)+\frac{2}{3}r_+r_-^2\right]}{4\pi r_+r_-(r_++r_-)\left(r_0^2-\frac{2}{3} r_-^2\right)}\,,
\eea
it is straightforward to check that, due to violation of $T_+S_+=T_-S_-$, the entropy product of the rotating Bardeen solution with small $r_0$ is not universal. In other words, adding rotation to the Bardeen and Hayward BHs does not lead to the universality of entropy product.

\subsection{The Frolov-Zelnikov BH}
In \cite{Frolov:2017rjz} an evaporating (time evolving) RBH with a de Sitter core is introduced. The metric is in the form
\bea \label{frolmetr}
ds^2&=&-f dt^2+\frac{dr^2}{f}+r^2 \left(d\theta^2+\sin^2 \theta d\phi^2\right),\nn\\
f&=&1-\frac{2M(t) r^2}{r^3+2M(t)\ell^2+\ell^3}\,,
\eea
in the static case, this solution differs from the Hayward BH  by an extra term $\ell^3$ in the denominator of function $f$. This term guarantees the metric smoothness at $r=0$ in the limit $M(t)\to 0$. For this solution, one can find the horizons by solving $f=0$ (in the static case $M(t)=M$) as
\bea
r_+&=&\frac{\ell}{2}+M+\frac12 \sqrt{4M(M-\ell)-3\ell^2},\\
r_-&=&\frac{\ell}{2}+M-\frac12 \sqrt{4M(M-\ell)-3\ell^2},\qquad r_0=-\ell\,.\nn
\eea
For Hayward and Frolov solutions $\ell$ is positive, so the third root ($r_0=-\ell$) is a negative valued radius, which is not relevant to our study. Now it is possible to find the entropy and temperature as
\bea
S_+&=&\frac{\pi}{2}(\ell+2M)\big(2M-\ell+\sqrt{4M(M-\ell)-3\ell^2}\big)\,,\nn\\
S_-&=&\frac{\pi}{2}(\ell+2M)\big(2M-\ell-\sqrt{4M(M-\ell)-3\ell^2}\big)\,,\nn\\
T_+&=&\frac{Mr_+\big(\frac{r_+^3}{2}-\ell^3-2M\ell^2\big)}{\pi (r_+^3+2\ell^2M+\ell^3)^2}\,,\nn\\
T_-&=&\frac{Mr_-\big(\frac{r_-^3}{2}-\ell^3-2M\ell^2\big)}{\pi (r_-^3+2\ell^2M+\ell^3)^2}\,,
\eea
it is easy to check that $T_+S_+\neq T_-S_-$ which means that the entropy product is mass dependent in the case of the Frolov-Zelnikov BH.

\subsection{Five dimensional RBHs}
In this part we investigate the entropy product for some five dimensional BHs. We consider regular solutions with both Minkowski and de Sitter cores.

\subsubsection{Static RBH}
Five dimensional static RBH is introduced in \cite{Ahmed:2020dzj}. The line element is in the form
\bea \label{stat5}
ds^2 \!&=&\!-f(r)dt^2 +\frac{dr^2}{f(r)}+r^2\big(d\theta^2\!+\sin^2\!\theta d\phi^2\!+\cos^2\! \theta d\psi^2\big),\nn\\
&& \hspace{2cm} f(r)=1-\frac{M }{r^2}e^{-k/ r^2}\,.
\eea 
This solution which satisfies the equations of motion of 5D Einstein gravity coupled to nonlinear electrodynamics, is a RBH with Minkowski core at $r=0$. Here $M$ is the black hole mass and $k$ is related to the magnetic monopole charge ($q$) via $q^2=Mk$. The solution (\ref{stat5}) reduces to 5D Schwarzschild-Tangherlini BH when $k=0$ and one can also find the Minkowski space-time by setting $M=0$. For this RBH temperature on the outer and inner horizons are \cite{Ahmed:2020dzj}
\be
T_+=\frac{1}{4\pi r_+}\left(1-\frac{k}{r_+^2}\right), \quad T_-=\frac{-1}{4\pi r_-}\left(1-\frac{k}{r_-^2}\right),
\ee
and the entropies are
\be
S_+=\frac{\pi^2}{2}r_+^3\,, \quad  S_-=\frac{\pi^2}{2}r_-^3.
\ee
This solution is not universal since $T_+S_+\neq T_-S_-$.

\subsubsection{Magnetically charged Myers-Perry solution}
Applying the Newman-Janis algorithm on the static solution of the previous subsection, a 5D rotating RBH is constructed in \cite{Ahmed:2020dzj}. This solution is a magnetically charged Myers-Perry BH with Minkowski core. The metric is 
	\bea \label{5dmpk}
	&ds^2&\!\!=- \left(1-\frac{\mathcal{M}}{\rho^2}\right) dt^2+\frac{r^2 \rho^2}{\Delta} dr^2 + \rho^2 d\theta^2 -\frac{2 \mathcal{M} a \sin^2 \theta }{\rho^2} dt d\phi - \frac{2 \mathcal{M} b \cos^2 \theta }{\rho^2} dt d\psi \\
	&+&\!\!\! \sin^2\! \theta \left(\! r^2 \!+ a^2 \!+\frac{\mathcal{M} a^2 \sin^2\! \theta }{\rho^2}\right)\! d\phi^2 
	+ \cos^2\! \theta \left(\! r^2 \!+ b^2 \!+\frac{\mathcal{M} b^2 \cos^2\! \theta }{\rho^2}\right)\! d \psi^2+ \frac{2 \mathcal{M} a b \sin^2\! \theta \cos^2\! \theta }{\rho^2} d\phi d\psi,\nn
	\eea
where $\rho$, $\mathcal{M}$ and  $\Delta$ are defined as
\be
\rho^2=r^2+a^2\cos^2 \theta +b^2\cos^2 \theta ,  \quad \mathcal{M}=M e^{-k/r^2}, \quad \Delta=(r^2+a^2)(r^2+b^2)- \mathcal{M}r^2.
\ee
The above solution which is a RBH with Minkowski core at $r=0$, is characterized by four parameters: mass $M$, rotation parameters $a,b$ and $k$ which is related to the magnetic charge. Setting $k=0$ in (\ref{5dmpk}) one can recover the 5D Myers-Perry BH, so $k$ can be interpreted as a deviation from the Myers-Perry geometry. Moreover by inserting $a=b=0$ the solution reduces to Schwarzschild-Tangherlini BH.

Although one may find horizons by solving $\Delta=0$, it is more convenient to find $M$ and $a$ in terms of $r_{\pm}$. For the regular solution with small charge parameter $k$, we calculate the entropy and temperature on the horizons. The result is
\bea
S_+&=&\frac{\pi^2(k-r_+^2)(b^2+r_+^2)(b^2+r_-^2)}{2r_+(k+b^2)}\,,\nn\\ S_-&=&\frac{\pi^2(k-r_-^2)(b^2+r_+^2)(b^2+r_-^2)}{2r_-(k+b^2)}\,,\nn
\eea
\bea
T_+&=&\frac{r_+^4(r_-^2-b^2-2k)-k^2(b^2+r_-^2)}{2\pi r_+^3(b^2+r_-^2)(k-r_+^2)}\,, \nn\\
T_-&=&\frac{r_-^4(r_+^2-b^2-2k)-k^2(b^2+r_+^2)}{2\pi r_-^3(b^2+r_+^2)(r_-^2-k)}\,.
\eea
Now it is straightforward to check that the condition $T_+S_+= T_-S_-$ does not hold and so the entropy product for the above solution is not universal.

\subsubsection{Electrically charged rotating Bardeen BH }
5D Bardeen BH with electric charge and its rotating version are introduced in \cite{Ali:2018boy,Amir:2020fpa}. The line element for electrically charged rotating Bardeen BH is \cite{Amir:2020fpa}
\bea \label{crB}
&&\!\!\!\!ds^2=-dt^2 +\frac{m(r)}{\rho^2} \left(dt-a\sin^2\theta d\phi + b\cos^2\theta d\psi \right)^2 \\
&&\!\!\!\!+\frac{r^2 \rho^2}{\Delta}dr^2\!+\rho^2 d\theta^2\! +(r^2 \!+\!a^2)\!\sin^2\!\theta d\phi^2\! +(r^2 \!+\!b^2)\!\cos^2\!\theta d\psi^2\!\!,\nn
\eea
where $\rho$, $m(r)$ and $\Delta$ are as the following
\bea
&&\!\!\!\rho^2=r^2\!+a^2\cos^2\!\theta+b^2\sin^2\!\theta, \quad m(r) = \mu \left(\!\!\frac{r^3}{r^3 \!+\! q_e^3}\!\right)^{\!4/3}\!,\nn\\
&&\!\!\!\Delta=\big(r^2+a^2\big)\left(r^2+b^2\right)-m(r)r^2.
\eea
The above solution is characterized by four parameters $a$, $b$, $\mu$ and $q_e$ where $a$, $b$ are rotation parameters around $\phi$, $\psi$ axis  and $\mu$ and $q_e$ are related to the mass and electric charge of the BH respectively. Setting $q_e=0$ in (\ref{crB}), the Myers-Perry BH is recovered so $q_e$ determines the deviation of charged rotating Bardeen BH from the Myers-Perry solution.

Horizons of this solution are the roots of $\Delta=0$ and due to the complicated form of the metric, one finds messy terms for the entropies and temperatures on the horizons. In the following, for the sake of simplicity, we consider single rotating Bardeen BH with small value of charge which is obtained by setting $b=0$ and doing Taylor expansion for $m(r)$. Noticing the ranges of angles in this solution as $\theta \in [0,\pi/2]$ and $\phi,\psi \in [0,2\pi]$, we calculate the entropy and temperature on the inner and outer horizons. The result is
\bea
S_+&=&\frac{\pi^2 r_+r_-^3(r_-+r_+)(4q_e^3-3r_+^3)}{8q_e^3(r_-^2+r_-r_++r_+^2)}\,,\nn\\ S_-&=&\frac{\pi^2 r_-r_+^3(r_-+r_+)(4q_e^3-3r_-^3)}{8q_e^3(r_-^2+r_-r_++r_+^2)}\,,\\
T_+&=&\frac{-3r_-^4-3r_-^3r_++2r_-^2r_+^2+2r_-r_+^3+2r_+^4}{3\pi r_-^3 r_+^4(r_-+r_+)}q_e^3,\nn\\
T_-&=&\frac{-3r_+^4-3r_+^3r_-+2r_-^2r_+^2+2r_+r_-^3+2r_-^4}{3\pi r_+^3 r_-^4(r_-+r_+)}q_e^3,\nn
\eea
one can easily check that $T_+S_+\neq T_-S_-$ which means the universality of entropy product fails in this solution.

\section{RBHs with universal entropy product} \label{s4}
Until now, we observed that the universality of entropy product fails for the RBHs. However there are some regular solutions which respect the universality.

\subsection{The Kerr-like RBH} \label{kerrlike}
In \cite{Ghosh:2014pba,Simpson:2021zfl} it has been introduced a regularized Kerr BH with modified mass $m \to m(r)=m e^{-\ell/r}$. The metric is given by 
\bea \label{mkerr}
ds^2 &=& - \frac{\Delta'}{\Sigma}(d t-a\sin^2\theta\,d\phi)^2+ \frac{\Sigma}{\Delta'}\,d r^2 + \Sigma\,d\theta^2 
+ \frac{\sin^2\theta}{\Sigma}[(r^2+a^2)\,d\phi-a\,d t]^2\,, \nn\\
\Sigma&=&r^2+a^2\cos^2\theta\,, \qquad \qquad \Delta'=r^2+a^2-2mr\,e^{-\ell/r}\,.
\eea
This RBH is characterized by its  mass $m$, rotation parameter $a$ and the regularization parameter $\ell$. Setting $\ell=0$, (\ref{mkerr}) reduces to the Kerr metric, so $\ell$ can be also viewed as the deviation form Kerr. This solution at $r \to \infty$ reaches the Kerr BH, while the ring singularity of Kerr is replaced by an asymptotically Minkowski spacetime \cite{Simpson:2021zfl}. For the small values of $\ell$, we keep the first order in Taylor expansion of the metric. We also rewrite $a$ and $m$ in terms of $r_{\pm}$. Now the entropy and temperature for this perturbed solution takes to the following form 
\bea \label{stmkerr}
S_+\!\!&=&\!\pi (r_+\!+\!r_-)(\ell\!-\!r_+), \quad S_-=\pi (r_+\!+\!r_-)(\ell\!-\!r_-), \\
T_+\!\!&=&\! \frac{r_--r_+}{4\pi(r_-\!+r_+)(\ell\!-r_+)}, \quad T_-= \frac{r_--r_+}{4\pi(r_-\!+r_+)(\ell\!-r_-)},\nn
\eea
it is easy to check that $T_+S_+= T_-S_-$ is satisfied, which means that the entropy product is universal. We also calculate the angular momentum for the solution with small $\ell$ as
\be \label{jpk}
J=\frac12 (r_++r_-)\,\sqrt{r_-r_+-\ell(r_++r_-)}=ma\,.
\ee
Noticing (\ref{stmkerr}), (\ref{jpk}) and by ignoring the $\ell^2$, $\ell^3$, ... terms, one finds the universality of entropy product as
\be \label{unipk}
S_+S_-=4\pi^2 J^2\,.
\ee
In the next section we will use this result to find the central charge of the dual CFT for the regular Kerr-like BH.

\subsection{Reissner-Nordstr\"{o}m outside a de Sitter core} \label{reis}
In \cite{Lemos:2011dq} a regular solution is constructed by connecting a de Sitter space to the core of Reissner-Nordstr\"{o}m BH. The metric is written in the static spherically symmetric form
\be
ds^2 = -B(r)dt^2 + A(r) dr^2 + r^2(d\theta^2+\sin^2 \theta d\phi^2)\,.
\ee
In order to remove the singularity at $r=0$, it is supposed a sphere with radius $r_0$, centered at $r=0$, which is filled by a static charged perfect fluid distribution with spherical symmetry. Solving the equations of motion, functions $A(r)$ and $B(r)$ are obtained as \cite{Lemos:2011dq}
\be
B(r) = A^{-1}(r)  = \left\{\begin{array}{l l}
	\displaystyle{ 1-\frac{r^2}{R^2}}, &  {r\leq r_0},\\
	\displaystyle{1- \frac{2m}{r} + \frac{q^2}{r^2}}, &  {r\geq r_0}.
\end{array} \right.
\ee
The above solution is determined by its mass ($m$), charge ($q$), radius of de Sitter space at the core ($R$) and radius of matter distribution ($r_0$). In the region $r>r_0$ the solution is given by the Reissner-Nordstr\"{o}m metric which has two horizons at $r_\pm=m\pm\sqrt{m^2-q^2}$. It has been shown in \cite{Lemos:2011dq} that $r_0\leq r_-$, in other words the region of matter distribution lies inside the inner horizon and the spacetime for $r \geq r_-$ is given by the Reissner-Nordstr\"{o}m metric. The entropy and temperature on the inner and outer horizons for this RBH is just like that of the Reissner-Nordstr\"{o}m
\be 
S_+= \pi r_+^2\,, \quad S_-=\pi r_-^2\,,\qquad T_+=\frac{r_+-r_-}{4\pi r_+^2}\,, \quad T_-=\frac{r_+-r_-}{4\pi r_-^2}\,.
\ee
The universality is true, since the condition $T_+S_+= T_-S_-$ is satisfied. Moreover the entropy product for this solution is mass independent
\be \label{unirn}
S_+S_-=\pi^2 q^4\,.
\ee

\subsection{The topological star} \label{tops}
The topological star is a solution which is obtained from dimensional reduction of a 5D solution in Einstein-Maxwell theory \cite{Bah:2020ogh}. The starting point is  the static spherically symmetric metric in five dimension with a magnetic flux $F$
\bea \label{metr2}
ds^2\!&=&\!-f_S(r)dt^2+f_B(r)dy^2+\frac{dr^2}{f_S(r)f_B(r)}+r^2d\theta^2 \nn\\
&+&r^2\sin^2\!\theta d\phi^2, \qquad F=P\sin\theta d\theta\wedge d\phi\,,
\eea
coordinate $y$ in the above, parametrizes a circle of perimeter $2\pi R_y$ and the functions $f_S(r)$,  $f_B(r)$ and $P$ are given by
\bea
&&f_B(r)=1-\frac{r_B}{r}\,, \qquad f_S(r)=1-\frac{r_S}{r}\,, \nn\\
&&P=\pm\frac{1}{\kappa_5^{2}}\sqrt{\frac{3\,r_S\,r_B}{2}}\,,
\eea
where $\kappa_5$ is the gravitational coupling. By Kaluza-Klein reduction along $y$, one finds the following solution
\bea
ds_5^2\!\!&=&\!\!e^{2\Phi}ds_4^2+e^{-4\Phi}dy^2,\qquad  e^{2\Phi}=f_B^{-\frac12}, \\
ds_4^2\!\!&=&\!\!f_B^{\frac12}\big[-\!f_Sdt^2\!+\frac{dr^2}{f_Bf_S}+r^2d\theta^2\! +r^2\sin^2\!\theta d\phi^2\big]\!.\nn
\eea

In the case of $r_B=0$ the above metric reduces to the Schwarzschild BH. Note also that the coefficient $f_B^{1/2}$ is imaginary for $r<r_B$ which means that the spacetime ends at $r=r_B$. In other words the singularity at $r=0$ is excluded from the four dimensional BH. There are two horizons (roots of the term $f_Bf_S$) for this solution at $r_B$ and $r_S$ with the entropy and temperature
\be
S_S=\frac{\pi r_S^2}{G}\,, \quad S_B=\frac{\pi r_B^2}{G}\,,\qquad T_S=\frac{1}{4\pi r_S}\,,\quad T_B=\frac{r_S}{4\pi r_B^2}\,.
\ee
The universality is held since the condition $T_+S_+=T_-S_-$ is respected. For this solution the  mass and magnetic charge take to the form
\be \label{mq}
M=\frac{2\pi}{\kappa_4^2}\left(2r_S+r_B\right), \qquad
Q_m=\frac{1}{\kappa_4}\sqrt{\frac32\,r_B\, r_S}\,,
\ee
where $\kappa_4$ is the 4D gravitational constant which is related to $\kappa_5$ as $\kappa_4=\frac{\kappa_5}{\sqrt{2\pi R_y}}$. One can check that the entropy product for this regular solution is mass independent as \footnote{Considering (\ref{mq}) and Noticing that in the case of Schwarzschild black hole ($r_B=0$) we find $r_S=2M$, it is obvious that $\kappa_4=\sqrt{8\pi}$. In addition, we rescale the magnetic charge as $Q_m\to 4\sqrt{\frac{\pi}{3}}\,Q_m$.}
\be \label{univ1}
S_S\,S_B=\pi^2 \,Q_m^4.
\ee

\subsection{BHs in Einstein gravity coupled to extended nonlinear electrodynamics} \label{egnl}
A physical source for the RBHs is the nonlinear electrodynamics \cite{Ayon-Beato:1998hmi,Ayon-Beato:2000mjt}. For the Einstein gravity coupled to the nonlinear electrodynamics
\be \label{act}
S=\frac{1}{16\pi}\int d^4x\sqrt{-g}\,[R-\mathcal{L}(\mathcal{F})]\,,
\ee 
where $F_{\mu \nu}=dA_\nu$ is the field strength tensor, $\mathcal{F}=F_{\mu \nu}F^{\mu\nu}$ and Lagrangian density $\mathcal{L}$ is a function of $\mathcal{F}$, it has been shown \cite{Fan:2016hvf} that one can construct variety classes of RBHs by choosing different forms of $\mathcal{L}(\mathcal{F})$. In the following we review three classes of these RBHs.
\subsubsection{Bardeen class}
Considering the Lagrangian density 
\be
\mathcal{L}=\frac{4\mu \big(\alpha \mathcal{F} \big)^{5/4}}{\alpha \big( 1+\sqrt{\alpha \mathcal{F}} \big)^{1+\mu/2}} \,,
\ee
where $\mu>0$ and $\alpha>0$ are some constants, the general form for the two-parameter (mass and magnetic charge) BH solutions is
\be \label{bardclas}
ds^2=-f dt^2+\frac{dr^2}{f}+r^2(d\theta^2+\sin^2\theta d\phi^2)\,,\qquad f=1-\frac{2M}{r}-\frac{2\alpha^{-1} q^3 r^{\mu-1}}{\big( r^2+q^2 \big)^{\mu/2}}\,,
\ee
in the above, $q$ is related to the black hole magnetic charge and the ADM mass can be read
off from the asymptotic behavior of the metric 
\be
f=1-\frac{2(M+\alpha^{-1}q^3)}{r}+... ~,
\ee
which implies that $M_{ADM}=M+M_{em}$ where $M_{em}=\alpha^{-1}q^3$. In other words, the BH mass receives contributions from the Schwarzschild mass $M$ and the nonlinear effects that denoted by $M_{em}$. Since by setting $M=0$ and $\mu=3$ the Bardeen BH (\ref{bardmetr}) is recovered, (\ref{bardclas}) is called "Bardeen class". It has been observed \cite{Fan:2016hvf} that for a solution with $M=0$ and $\mu=2$ the entropy product is universal as \be \label{unibc}
S_+S_-=\pi^2 q^4\,.
\ee
In the next section we will use the above result to find the central charge of dual CFT.

\subsubsection{Hayward class}
The nonsingular BHs of the Hayward class are solutions to Einstein gravity coupled to a nonlinear electrodynamics with Lagrangian density \cite{Fan:2016hvf}
\be \mathcal{L}=\frac{4\mu(\alpha \mathcal{F} )^{(\mu+3)/4}}
{\alpha\big[1+(\alpha \mathcal{F})^{\mu/4} \big]^2} \,,
\ee
similar to the Bardeen class, for these BHs the metric is in the static spherically symmetric form (\ref{bardclas}) with function 
\be
f=1-\frac{2M}{r}-\frac{2\alpha^{-1} q^3\, r^{\mu-1}}{ r^\mu+q^\mu}\,.
\ee
Setting $M=0$ and $\mu=3$ one finds the Hayward solution, for this reason they called  "Hayward class". It has been found \cite{Fan:2016hvf} that  in the case of vanishing Schwarzschild mass ($M=0$) and $\mu=2$ the entropy product is universal as\be \label{unihc}
S_+S_-=\pi^2 q^4\,.
\ee
\subsubsection{new class}
By choosing the Lagrangian density as \cite{Fan:2016hvf},
\be \mathcal{L}=\frac{4\mu \mathcal{F}}
{\big[1+(\alpha \mathcal{F} )^{1/4} \big]^{\mu+1}}\,,
\ee
a "new class" of RBHs can be obtained. The metric is given by (\ref{bardclas}) with new function 
\be
f=1-\frac{2M}{r}-\frac{2\alpha^{-1} q^3 r^{\mu-1}}{(r+q)^\mu}\,.
\ee
For a solution with $M=0$ and $\mu=2$ the entropy product is universal as (\ref{unihc}). 

\section{Central charges and entropy product} \label{s5}

In sections \ref{s3} and \ref{s4} we studied the universality of entropy product in the case of RBHs. We found that universality is true for some of them and it fails for some others. According to  Thermodynamics method (reviewed in section \ref{s2}) when the entropy product is universal, one can easily read the central charge of the dual CFT from (\ref{unicc}). In the following, for RBHs in section \ref{s4} which respect the universality, we find central charges form (\ref{unicc}). Then we compare this result with the central charge calculated from the asymptotic symmetry group (ASG) method \cite{BH,Guica:2008mu}, to check the validity of Thermodynamics method.

\subsection{Central charge of The Kerr-like RBH}
In subsection (\ref{kerrlike}) we observed that for the small deviation parameter $\ell$, entropy product of the Kerr-like RBHs is universal as 
\be
S_+S_-=4\pi^2J^2,
\ee
Now using (\ref{unicc}), it is easy to read off the central charge from this entropy product
\bea \label{cc1}
c&=&\frac{6}{4\pi^2}\,\frac{\partial \big(4\pi^2J^2 \big)}{\partial J}=12 J\nn\\ &=&6(r_++r_-)\sqrt{r_-r_+-\ell(r_++r_-)},
\eea
where in the second line we use (\ref{jpk}). We are going to verify the above result by doing the Kerr/CFT analysis \cite{Guica:2008mu}. The first step is to find the near horizon metric of the extremal solution. Remember that we have written parameters $a$ and $m$ in the Kerr-like BH (\ref{mkerr}) in terms of $r_{\pm}$, so one obtains the extremal solution by setting $r_-=r_+$. Following the procedure as in \cite{Chen:2012yd,Golchin:2020mkh}, we find the near horizon extremal metric. The result is
\be \label{NHE1}
ds^2=\alpha(\theta)\big[-r^2dt^2+\frac{dr^2}{r^2}\big]+\beta(\theta)d\theta^2 +\gamma(\theta)(d\phi+f^\phi r dt)^2\!,
\ee
where $f^{\phi}$ and functions $\alpha(\theta)$, $\beta(\theta)$ and $\gamma(\theta)$ are
\bea
&&f^\phi=\frac{r_+(r_+-2\ell)}{\ell-r_+}\,, \quad \gamma(\theta)=\frac{4r_+(\ell-r_+)^2\sin^2\theta}{r_+-(2\ell-r_+)\cos^2\theta}\,,\nn\\
&&\alpha(\theta)=\beta(\theta)= -2r_+\left[\big(\ell-\frac{r_+}{2}\big)\cos^2\theta-\frac{r_+}{2}\right]\,.
\eea
By choosing the adequate boundary conditions given in \cite{Guica:2008mu} and doing the calculations, we find the central charge as
\be \label{cc2}
c=\frac{3}{2\pi}f^\phi\!\!\int \!\!\! d\theta d\phi \sqrt{\beta(\theta) \gamma(\theta)}=12 \sqrt{r_+^3(r_+\!-2\ell)}=12J_{\!ext},
\ee
where $J_{ext}$ is the angular momentum of the extremal solution. To compare the central charges (\ref{cc1}) and (\ref{cc2}) it is important to note that in the Kerr/CFT analysis we have inserted the extremality on the solution. Now applying the extremality condition $r_-=r_+$ on (\ref{cc1}), it is obvious that the result is in complete agreement with (\ref{cc2}) which means the validity of Thermodynamics method.
It is worth mentioning that using the Thermodynamics method, one finds central charge for the generic (non-extremal) solution while in the Kerr/CFT analysis, the central charge is obtained for the extremal solution.

\subsection{Central charge of the charged RBHs}
For the regular Reissner-Nordstr\"{o}m and  topological star in \ref{reis} and \ref{tops} we found that the entropy product is universal\be 
S_+S_-=\pi^2 q^4\,,
\ee
where $q$ is the charge (electric or magnetic) of the solution. The same relation is obtained in the case of RBHs of Einstein gravity coupled to nonlinear electrodynamics (subsection \ref{egnl}). Now it is easy to find the central charge of the dual CFT using Thermodynamics method (\ref{unicc}) as
\be 
c=\frac{6}{4\pi^2}\,\frac{\partial \big(\pi^2q^4 \big)}{\partial q}=6q^3\,.
\ee
One can also find central charges for charged solutions using the ASG formalism. In fact this is done in \cite{Garousi:2009zx}, the result is 
\be
c=6q^3,
\ee
which is in agreement with the result of Thermodynamics method.

\section{Conclusions} 
In order to extend our intuition about the entropy product law and its relation to dual CFTs of the BHs, we investigated the entropy product for RBHs. We considered variety types of regular solutions including the rotating Bardeen and Hayward BHs, the Frolov-Zelnikov solution, the Kerr-like RBH, the regular Reissner-Nordstr\"{o}m solution, topological star, 5D charged  rotating RBHs and regular solutions in extended nonlinear electrodynamics.

It seems that there is no rule for the RBHs that makes their entropy product universal. In other words, regardless of the charges or asymptotic behavior at $r\to 0$ limit or even the theory that RBH satisfies its equations of motion, the universality of entropy product is true for some RBHs but fails for some others!
The similar pattern is observed for the singular BHs. For instance, the universality is true for the Kerr BH and 5D Myers-Perry but it fails for the Myers-Perry in $D\geq 6$. The same situation is for the BTZ (which is universal) and Kerr-AdS in $D\geq 4$ (non-universal) BHs. One can deduce from these observations that in generic, the universality is a characteristic for the ``solutions" and not for the ``theories" \footnote{There is an exception here \cite{MahdavianYekta:2020lde}: In the case of theories in which the action contains parity violating terms (such as the Chern-Simons), one finds that  $c_R\neq c_L$ which means $T_+S_+\neq T_-S_-$.}. 

So what is the advantage of entropy product universality? One may find answer in the relation between entropy product and dual CFTs of the BHs. It has been discussed \cite{Chen:2013rb} that when the entropy product is universal, the central charges of the dual CFTs can be easily read off from the entropy product according to Thermodynamics method (\ref{unicc}). 

In the case of RBHs with universal entropy product, we found the central charges using the Thermodynamics method. We then found the central charges from the ASG method. These two results are the same which shows the validity of Thermodynamics method for RBHs. In other words, universality of the entropy product provides an easy way to find central charges of the dual CFTs. It is worth to mention that, in the ASG method one takes the extremal limit of the solutions and so the central charges are obtained for the extremal BHs. While the Thermodynamics method yields central charges of the generic (non-extremal) BH solutions. 

\section*{Acknowledgment}
I would like to thank IPM, school of Physics, because of useful discussions and hospitality during my visit. I would also like to thank S. Sadeghian for reading the manuscript and her valuable comments.



\end{document}